\documentclass{article}
\usepackage{spconf,amsmath,graphicx,hyperref}
\usepackage[font={footnotesize},labelfont={bf}]{caption}
\usepackage[english]{babel}
\usepackage[autostyle, english = american]{csquotes}
\MakeOuterQuote{"}

\makeatletter
 \renewcommand\section{\@startsection {section}{1}{\z@}%
     {-3ex \@plus -1ex \@minus -.2ex}%
     {1.3ex \@plus.2ex}%
    {\centering\bfseries}}

\usepackage{background}
\SetBgAngle{0}
\SetBgScale{1}
\SetBgColor{black}
\SetBgOpacity{1}
\SetBgPosition{.51\paperwidth,-1.08\textheight}
\SetBgContents{
{\parbox{\paperwidth}{%
\footnotesize \copyright 2025 IEEE.  Personal use of this material is permitted. 
Permission from IEEE must be obtained for all other uses, in any current or future media,\\
including reprinting/republishing this material for advertising or 
promotional purposes, creating new collective works, for resale or redistribution\\
to servers or lists, or reuse of any copyrighted component of this work in other works. 
\mbox{}\\
     \mbox{}\\}
     }%
}

\title{MUSHRA--1S: A scalable and sensitive test approach\\for evaluating top-tier speech processing systems}
\name{Laura Lechler and Ivana Balić}
\address{Collaboration AI, Cisco Systems, United Kingdom}
\begin{document}
%
\maketitle
\begin{abstract}
Evaluating state-of-the-art speech systems
necessitates scalable and sensitive evaluation methods to detect subtle but unacceptable artifacts. Standard MUSHRA is sensitive but lacks scalability, while ACR scales well but loses sensitivity and saturates at a high quality. To address this, we introduce MUSHRA--1S, a single-stimulus variant that rates one system at a time against a fixed anchor and reference. Across our experiments, MUSHRA--1S matches standard MUSHRA more closely than ACR, including in the high-quality regime, where ACR saturates. MUSHRA--1S also effectively identifies specific deviations and reduces range-equalizing biases by fixing context. Overall, MUSHRA--1S combines MUSHRA--level sensitivity with ACR--like scalability, making it a robust and scalable solution for benchmarking top-tier speech processing systems.
\end{abstract}
\section{Introduction}
\label{sec:intro}
Recent advances in deep learning for speech processing have enabled high speech quality with fewer resources. For example, audio codecs now perform very well at low bitrates, e.g., \cite{soundstream2022, kumar_high-fidelity_2023}, and speech enhancement models can reconstruct speech from very low SNR cases \cite{richter_2023}. Generative neural networks have recently succeeded in speech coding, enhancement, bandwidth expansion, and text-to-speech synthesis. However, these can generate high-quality human-like speech that may differ from the original input in unacceptable ways, such as changes in voice characteristics, phoneme errors, or inappropriate prosody. Hence, evaluation should be reference-based, scalable, and highly sensitive so it can quantify small quality differences as well as reference-relative errors at near-transparent quality.


Standard MUltiple Stimuli with Hidden Reference and Anchor (MUSHRA) \cite{itu_itu-r_2015} framework is effective and sensitive, including when conducted via crowdsourcing \cite{lechler_crowdsourcing_2025}, but it does not scale to many systems. ITU-R BS.1534-3 recommends at most 12 stimuli (ideally 7) for lab-based tests \cite{itu_itu-r_2015}, and in crowdsourcing we find 5--6 is optimal for avoiding listener fatigue \cite{lechler_crowdsourcing_2025}. As a result, MUSHRA is impractical for benchmarking many systems or tracking long-term performance.

Absolute category ratings (ACR) \cite{itu_p800_1998}, which yield mean opinion scores (MOS), are often preferred for their scalability, cost-effectiveness, and simplicity. However, four key limitations must be considered:
\begin{enumerate}
\itemsep-0.4em
\item \textbf{Absolute category ratings are not truly absolute.} Studies show range-equalizing biases \cite{zielinski_biases_2008, le_maguer_back_2022, cooper_investigating_2023}, making careful condition selection crucial.
\item \textbf{The 5-point scale limits sensitivity.} Its coarseness and lack of reference prevent raters from penalizing small degradations \cite{raake_how_2010, koster_comparison_2015, perrotin_refining_2025}.
\item \textbf{Scale extremes saturate easily.} The polar labels "bad" and "excellent" cover a wider perceptual range than intermediate levels \cite{koster_comparison_2015}.
\item \textbf{High-quality but incorrect (hallucinated) outputs go undetected without a reference.} Reference files are needed to catch high-quality generative errors \cite{lechler_crowdsourcing_2025}.
\item \textbf{Listener expectations shift over time.} What is rated "excellent" changes with exposure to state-of-the-art systems \cite{le_maguer_back_2022}, particularly for synthetic speech.
\end{enumerate}
Recommendations to mitigate these issues include adding implicit anchors and references \cite{perrotin_refining_2025}, careful listener training, extended continuous scales \cite{koster_comparison_2015}, or a staggered approach combining ACR and MUSHRA for top systems \cite{perrotin_refining_2025}.
To balance scalability and reference-based evaluation, Müller et al.\ \cite{muller_speech_2024} recommend degradation category ratings (DCR) for neural audio codecs, noting lower sensitivity than MUSHRA and scale saturation at high quality. The DCR scale may also be less adaptable for specific tasks.

Given these findings, we see a need for a scalable, stable, and sensitive test for high-performing speech models. We therefore propose MUSHRA--1S, a MUSHRA-like test evaluating one system at a time with a stable anchor and reference. This approach compromises between MUSHRA's sensitivity and ACR's scalability. 
MUSHRA scores may also map to the ACR-MOS scale \cite{raake_how_2010}, complementing coarse ratings and aiding MOS predictor training. MUSHRA--1S could also reduce range-equalizing biases present in standard MUSHRA, as results become less dependent on the performance of other systems included in the test. While the direct comparison is lost, the range-equalizing bias may be limited to the constant reference and anchor context.

\section{Experiment Setup}
\label{sec:exp_setup}
\subsection{General test procedure}
\label{gen_procedure}
The data set comprises 100 originally full-band (44.1--48 kHz) high-quality clean speech files and was designed to ensure gender balance, accent diversity, and inclusion of both adult and child speech. The set served as the clean blind test set in the Low-Resource Audio Codec (LRAC) Challenge 2025\footnote{https://crowdsourcing.cisco.com/lrac-challenge/2025/} and is publicly available\footnote{https://github.com/cisco/multilingual-speech-testing/tree/main/LRAC-2025-test-data/blind-test-set/track-1/clean/}.

All tests were conducted on Prolific, with compensation based on an hourly rate of \$8.50. Participants were recruited using the following screener criteria: first language English, no hearing difficulties, no cochlear implant, 98\%+ approval rate, and 110+ studies approved. Each condition was first tested as a MUSHRA--1S, followed by a standard MUSHRA test with all conditions, and then as separate ACR tests. MUSHRA and MUSHRA--1S used the survey setup tools and procedure presented in \cite{lechler_crowdsourcing_2025}. In line with the Preliminary Study \ref{preliminary_exp}, we collected a minimum of 6 responses per file. ACR studies followed P.808, with two gold questions (one low-quality, one high-quality) and two catch trials randomly added to 20 rating questions. Catch trials were instructional checks; gold questions included an Opus 6 kbps sample and a 24 kHz reference. Eight responses per file were collected, the minimum recommended in \cite{naderi_open_2020}.

For normalization, mean scores for each condition in the standard MUSHRA test established anchor and reference targets for normalizing MUSHRA--1S per-file mean scores. ACR scores were not normalized directly; instead, ACR and MUSHRA-like MOS were plotted on separate y-axes and approximately aligned using adjusted y-axis limits. In long-term monitoring use cases, the scores can be adjusted to constant values for the anchor and reference, e.g., values obtained in the initial round of testing or 0--100 normalization. 
As MUSHRA tests are standardized and widely accepted as accurate and reliable, we consider the standard MUSHRA tests the ground truth test that MUSHRA--1S and ACR are compared to.

\subsection{Preliminary Study: Number of responses per file}
\label{preliminary_exp}
In a preliminary study, we tested how many responses per file are required to produce a stable mean score with acceptable confidence intervals across the 100-file test set. 
The codec tested in this experiment is an experimental internal codec at 1 kbps and shows a wide range of output qualities. 
We assume that many other codecs will produce more homogeneous outputs and that the number established on this more variable test case will ensure sufficient accuracy on other codecs in general. Approximately 30 responses were collected for each of the 100 files in the test set. 

\subsection{Experiment 1: Medium-to-high quality codecs}
\label{mid_qual_exp}
To evaluate the general performance and consistency of the MUSHRA--1S approach with respect to a standardized MUSHRA test and absolute category ratings (ACR), Exp.~1a compared three different flavors of the well-known codec Opus \cite{valin_definition_2012} (version 1.5.2) at 9 kbps: Opus (basic complexity) was compared with the enhancement options LACE \cite{buthe_lace_2023} and NoLACE \cite{buthe_nolace_2024}. Opus at 6 kbps was chosen as the anchor condition. Clean wide-band (WB) files at 16 kHz sampling rate (Fs) served as reference files. 
We considered MUSHRA scores in the range of [0, 50] as low, [50, 75] as medium, and [75, 100] as high quality in tests with Opus 6 kbps anchor.

A second set of tests (Exp.~1b) examined how the test accuracy of the three test approaches is affected by selecting a higher-quality anchor (Opus 9 kbps). The test setup "zooms into" a slightly higher quality range comprising four different internal WB) codecs (Models A--D) in their development stage. The reference condition was also clean WB.

\subsection{Experiment 2: Effect of anchor quality}
\label{high_qual_exp_low_anchor}
We further investigated the sensitivity of the three test methods (MUSHRA--1S, MUSHRA, and ACR) when evaluating high-quality codecs. We selected four high-performing codecs within a narrow range on the high end of the MUSHRA--1S scale: Stablecodec \cite{parker_scaling_2025} 1 kbps, Bigcodec \cite{xin_bigcodec_2024} 1 kbps, ESC \cite{gu_esc_2024} 6 kbps, and DAC \cite{kumar_high-fidelity_2023} 8 kbps. All of these codecs either natively produced WB outputs (16 kHz Fs) or, in the case of DAC, stimuli were downsampled to the same Fs. Clean reference files at 24 kHz Fs were used to align with the tests reported in \cite{nc_bm} and to examine test sensitivity in the very high quality range. 
In this controlled experiment, the effect of different quality anchors was further investigated. For Exp.~2a, the anchor condition was Opus WB at 6 kbps. For Exp.~2b, a higher-quality anchor (Opus WB 9 kbps) was chosen to see if this further increased the test sensitivity of the MUSHRA-like tests. 

\section{Results and Discussion}
\label{sec:results}

\subsection{Preliminary Study: Number of responses per file}

\begin{figure}[t!]
    \centering
    \includegraphics[width=1.0\linewidth]{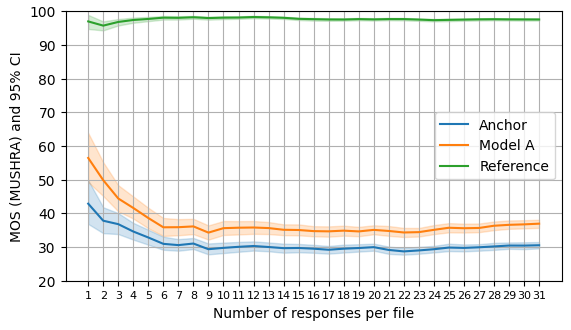}
    \vspace{-7mm}
    \caption{Mean scores and 95\% confidence intervals computed across 100 test files with increasing numbers of responses per file.}
    \label{fig:prelim_results}
    \vspace{-4mm}
\end{figure}
From the progression of mean scores and the 95\% confidence interval (CI), illustrated in Figure \ref{fig:prelim_results}, we identified that the overall mean scores for all files in the test set for Anchor and Model A settled at approximately 6 responses per file. Although the 95\% CI can be seen to further decrease, and small changes in the mean can occur (e.g., at 9 responses per file), we consider this precision sufficient for the minimum number of responses per file. In studies such as \cite{nc_bm}, resource constraints are of importance to consider, and slight imprecisions may be acceptable when many conditions need to be evaluated. For studies that require higher precision and a narrower CI, our preliminary study suggests that approximately 15 responses per file may be desirable.

\subsection{Experiment 1: Medium-to-high quality codecs}
\begin{figure}[b!]
    \centering
    \vspace{-6mm}
    \includegraphics[width=1.0\linewidth]{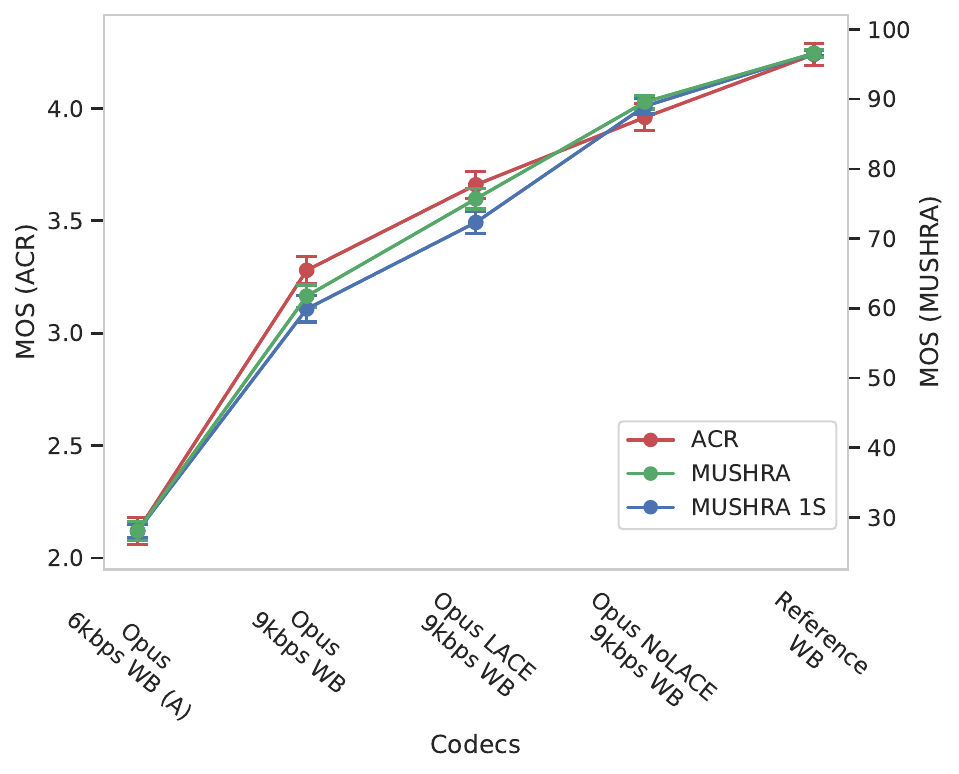}
    \vspace{-7mm}
    \caption{Exp.~1a: Test results for three well-known Opus 9 kbps versions spanning a wide quality range in the presence of a lower anchor.}
    \label{fig:mid_low_anchor}
\end{figure}

\subsubsection{Exp.~1a: Wide quality range with low anchor}
The overall mean scores and 95\% CIs obtained for Exp.~1a are presented in Figure \ref{fig:mid_low_anchor}. All three test types (standard MUSHRA, MUSHRA--1S, and ACR) achieved the same ranking in the experiment with three well-distinguishable conditions and are generally in good agreement with each other. For the MUSHRA-like tests, we observed a maximum absolute overall score difference of 3.38 points on the Opus LACE condition. For the ACR test, we observed a slightly flatter curve for the codecs tested. This may indicate slightly lower test sensitivity than for the MUSHRA-like tests. Although the effect is expected, due to the coarser rating scale and the absence of an explicit anchor and reference, it was small. We hypothesize that in this test, the well-differentiated test conditions, training exposure to the expected range of qualities, and use of gold questions consisting of one example of both anchor and reference, provided enough context to encourage raters to use the rating scale in a fairly consistent way, in line with previous research \cite{perrotin_refining_2025}. 
ANOVA tests with normalized test scores were conducted to compare the mean scores for test conditions across test types based on file means. A \textit{p}-value $\le$0.05 was considered a significant difference. The test confirmed a significant difference between the MUSHRA and the MUSHRA--1S scores only for Opus LACE, while significant differences were found for all test conditions when comparing ACR and MUSHRA scores. This indicates better alignment between MUSHRA and MUSHRA--1S than between MUSHRA and ACR. 

\begin{figure}[b!]
    \centering
    \vspace{-6mm}
    \includegraphics[width=1.0\linewidth]{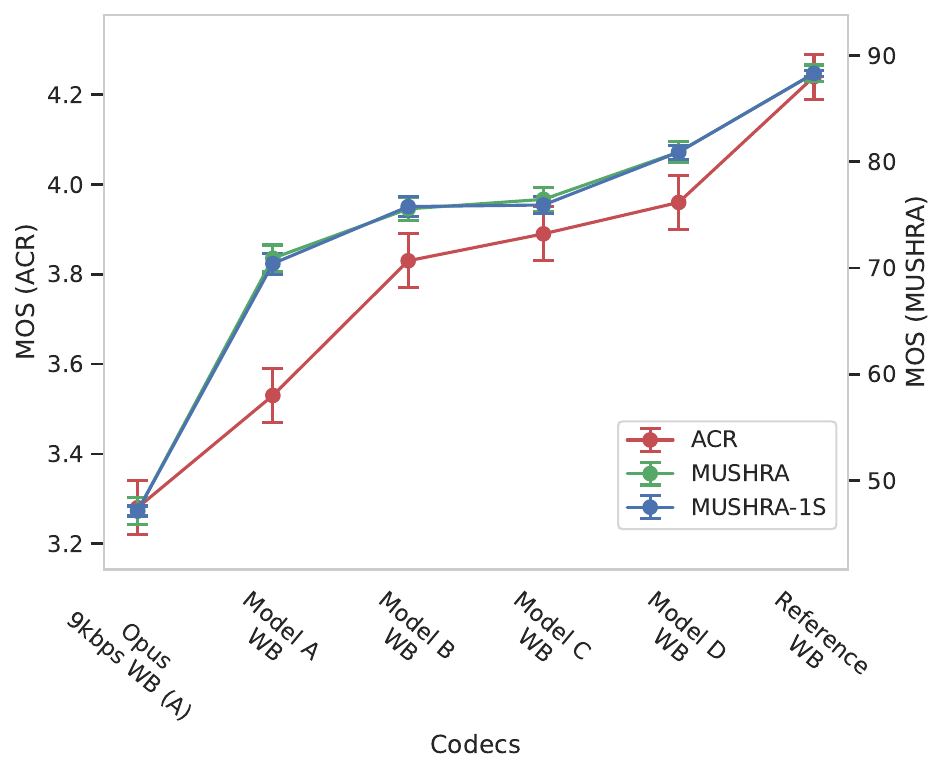}
    \vspace{-7mm}
    \caption{Exp.~1b: Test results for four internal codecs containing two very similar conditions in the presence of a higher anchor.}
    \label{fig:mid_high_anchor}
\end{figure}
\subsubsection{Exp.~1b: Narrower quality range with higher anchor}
Figure \ref{fig:mid_high_anchor} shows the overall scores for Exp.~1b, comparing four mid-to-high quality WB systems under a higher-quality anchor condition than in Exp.~1a. Visually, very good agreement between MUSHRA and MUSHRA--1S overall scores was observed. The largest absolute difference in overall scores between MUSHRA and MUSHRA--1S was 0.53 points. Although the general ranking of the ACR test is approximately in line with both MUSHRA-like tests, the relative distances between conditions (particularly, Anchor vs. Model A vs. Model B) are disproportionately reflected.

The discriminative capabilities of the test types between conditions, were examined through \textit{t}--tests (significance level \textit{p}$\le$0.05) between conditions within each of the test types. The distinctions between model pairs A--B, A--C, A--D, and B--D, as well as the distinctions of all models to both anchor and reference, were found significant in all three subjective test types. The difference between Models B and C was not found significant in either test type. However, the difference between Models C and D was significant in standard MUSHRA and MUSHRA--1S results, but not in ACR results, indicating slightly lower test sensitivity of ACR. 
This fine-grained replication of the MUSHRA results in the zoomed-in quality space of approximately the top half of the MUSHRA scale could not be achieved with the ACR test. 
Additionally, much larger 95\% CIs were observed in the ACR test compared to the MUSHRA-like tests, considering a normalized score space. 
The ANOVA tests comparing inter-test means revealed significant differences between ACR and MUSHRA scores for all test conditions. For MUSHRA--1S scores, however, significant differences were only observed for Model B, again, indicating better alignment between both MUSHRA-type tests than between ACR and MUSHRA.

\subsection{Experiment 2: Effect of anchor quality}
\begin{figure}[t!]
    \centering
    \includegraphics[width=1.0\linewidth]{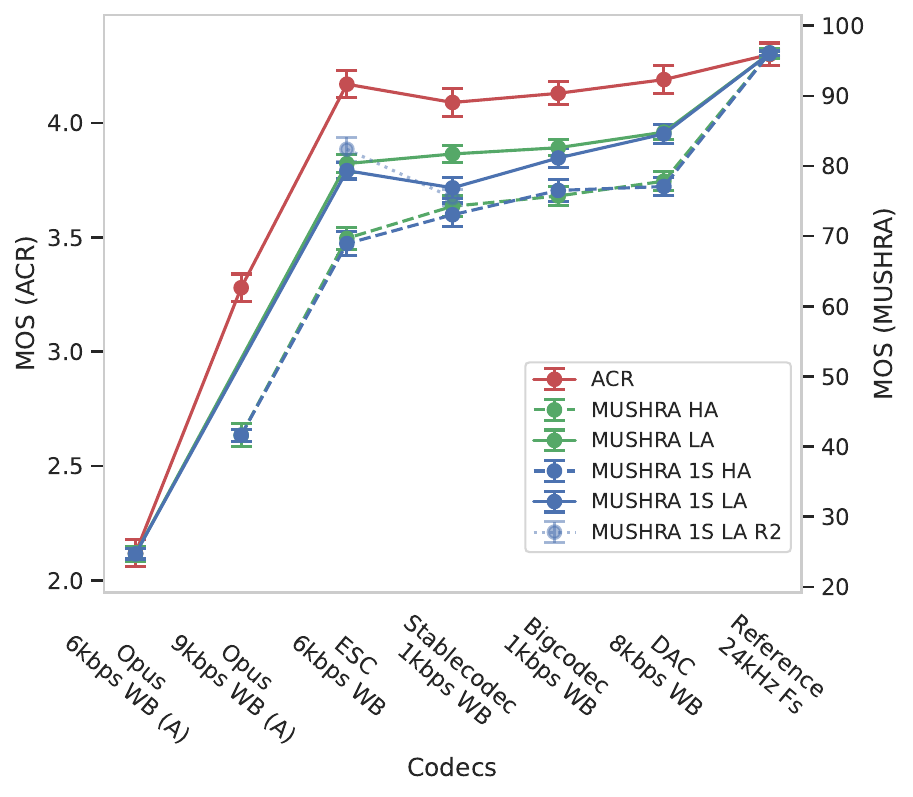}
    \vspace{-7mm}
    \caption{Exp.~2: Test results for four high-quality codecs assessed with either a lower anchor (LA; Exp.~2a)) or a higher anchor (HA; Exp.~2b)). MUSHRA--1S Run 2 (R2) results of ESC and Stablecodec are also shown.}

    \label{fig:MUSHRA_results_B_merged}
    \vspace{-4mm}
\end{figure}
\subsubsection{Exp.~2a: High-quality codecs with lower anchor (LA)}
Results for Experiment 2 are shown in Figure \ref{fig:MUSHRA_results_B_merged}. With a lower-quality anchor (Opus 6 kbps), overall MUSHRA--1S scores for Anchor, ESC, Bigcodec, DAC, and the 24 kHz Reference visually closely matched standard MUSHRA results (maximum absolute difference: 1.4 MUSHRA points). ANOVA tests comparing rescaled mean scores for a given condition between the different test types confirmed significant differences from MUSHRA scores for all conditions except Bigcodec in the MUSHRA--1S test and ESC in the ACR test. Therefore, neither of the indirect comparison methods (MUSHRA--1S and ACR) achieved good statistical alignment with MUSHRA, despite the visually well-aligned mean scores for MUSHRA and MUSHRA--1S. 

The poorer alignment was mainly caused by the lower scores for Stablecodec, relative to ESC and Bigcodec, in both MUSHRA--1S and ACR tests, but not MUSHRA. After ruling out factors like responses per file, outlier raters, locale, and test setup, the number of responses in the reference MUSHRA test was increased. We hypothesized that a direct comparison of four similar-quality codecs may pose a higher cognitive effort and may require more responses than single-condition tests. The pattern did not change with additional responses. MUSHRA--1S tests for ESC and Stablecodec were rerun (R2), confirming Stablecodec's lower score versus ESC. 

\subsubsection{Exp.~2b: High-quality codecs with higher anchor (HA)}
In Exp.~2b, with a higher-quality anchor (Opus 9 kbps), Stablecodec did not score lower than ESC in MUSHRA or MUSHRA--1S; in fact, it was rated higher, aligning with the standard MUSHRA results from Exp.~2a.
All other conditions being constant, we suspect the specific degradations of the lower anchor (Opus 6 kbps) in 2a caused the observed effect, but the pattern's presence in ACR and absence in 2b remain unexplained, requiring further study.

Under the presence of a higher anchor, MUSHRA--1S and MUSHRA scores aligned excellently. The narrower evaluation frame from the higher-quality anchor and reference lowered codec scores and accentuated the curve, especially among ESC, Stablecodec, and Bigcodec, likely due to reduced contrast with the anchor, allowing finer discrimination. However, at the high end, discriminative power decreased, with smaller differences between Bigcodec and DAC, likely due to the stark contrast with the 24 kHz reference. ACR tests, meanwhile, could not distinguish between adjacent codecs (except anchor and reference), indicating saturation.
In this test setup, no significant differences were observed in the ANOVA test comparing MUSHRA--1S with standard MUSHRA scores. ACR scores, on the other hand, significantly differed from MUSHRA scores for all conditions except DAC. This confirms an increased measurement precision of MUSHRA-type tests with a higher anchor.

\section{Conclusion}
\label{sec:conclusion}
The proposed MUSHRA--1S test showed better alignment of mean scores with standard MUSHRA tests than ACR, with both MUSHRA methods demonstrating high accuracy and sensitivity in the higher quality range, while ACR scores saturated for high-quality systems. Our experiments demonstrate that MUSHRA--1S reliably replicates MUSHRA scores, enabling accurate testing and unlimited scalability. Thus, we recommend MUSHRA--1S for benchmarking and long-term tracking studies, especially for evaluating very similar or high-quality systems, where it outperforms ACR.

Our findings also highlight the importance of carefully selecting anchors and references: closely bounding the expected quality range maximizes MUSHRA and MUSHRA--1S accuracy and sensitivity. We believe transformed MUSHRA--1S-MOS scores would complement existing data sets of ACR-MOS scores for training MOS predictor metrics and expand their effectiveness into the high-quality range, since MUSHRA methods are not limited by the saturation effects seen in ACR.

\vfill\pagebreak

\bibliographystyle{IEEEbib}
\bibliography{strings,MUSHRA-1s}

\begin{thebibliography}{10}

\bibitem{soundstream2022}
Neil Zeghidour, Alejandro Luebs, Ahmed Omran, Jan Skoglund, and Marco Tagliasacchi,
\newblock ``Soundstream: An end-to-end neural audio codec,''
\newblock {\em IEEE/ACM Transactions on Audio, Speech, and Language Processing}, vol. 30, pp. 495--507, 2022.

\bibitem{kumar_high-fidelity_2023}
Rithesh Kumar, Prem Seetharaman, Alejandro Luebs, Ishaan Kumar, and Kundan Kumar,
\newblock ``High-fidelity audio compression with improved {RVQGAN},''
\newblock in {\em Advances in Neural Information Processing Systems}, A.~Oh, T.~Naumann, A.~Globerson, K.~Saenko, M.~Hardt, and S.~Levine, Eds., 2023, vol.~36, pp. 27980--27993.

\bibitem{richter_2023}
Julius Richter, Simon Welker, Jean-Marie Lemercier, Bunlong Lay, and Timo Gerkmann,
\newblock ``Speech enhancement and dereverberation with diffusion-based generative models,''
\newblock {\em IEEE/ACM Transactions on Audio, Speech, and Language Processing}, vol. 31, pp. 2351--2364, 2023.

\bibitem{itu_itu-r_2015}
{ITU},
\newblock ``{ITU-R BS}.1534-3: Method for the subjective assessment of intermediate quality level of audio systems,'' 2015.

\bibitem{lechler_crowdsourcing_2025}
Laura Lechler, Chamran Moradi, and Ivana Balic,
\newblock ``Crowdsourcing {MUSHRA} tests in the age of generative speech technologies: A comparative analysis of subjective and objective testing methods,''
\newblock in {\em Interspeech 2025}, 2025, pp. 3160--3164.

\bibitem{itu_p800_1998}
{ITU},
\newblock ``P.800 : Methods for subjective determination of transmission quality,'' 1998.

\bibitem{zielinski_biases_2008}
Slawomir Zielinski, Francis Rumsey, and Søren Bech,
\newblock ``On some biases encountered in modern audio quality listening tests-a review,''
\newblock {\em Journal of the Audio Engineering Society}, vol. 56, no. 6, pp. 427--451, 2008.

\bibitem{le_maguer_back_2022}
Sébastien Le~Maguer, Simon King, and Naomi Harte,
\newblock ``Back to the future: Extending the blizzard challenge 2013,''
\newblock in {\em Interspeech 2022}, 2022, pp. 2378--2382.

\bibitem{cooper_investigating_2023}
Erica Cooper and Junichi Yamagishi,
\newblock ``Investigating range-equalizing bias in mean opinion score ratings of synthesized speech,''
\newblock in {\em Interspeech 2023}, 2023, pp. 1104--1108.

\bibitem{raake_how_2010}
Alexander Raake, Marcel Wältermann, Falk Schiffner, Ulf Wüstenhagen, and Bernhard Feiten,
\newblock ``How to talk about speech and audio quality with speech and audio people?,''
\newblock in {\em 3rd International Workshop on Perceptual Quality of Systems ({PQS} 2010)}, 2010, pp. 71--76.

\bibitem{koster_comparison_2015}
Friedemann Köster, Dennis Guse, Marcel Wältermann, and Sebastian Möller,
\newblock ``Comparison between the discrete {ACR} scale and an extended continuous scale for the quality assessment of transmitted speech,''
\newblock {\em Fortschritte der Akustik, {DAGA}}, vol. 3, pp. 150--153, 2015.

\bibitem{perrotin_refining_2025}
Olivier Perrotin, Brooke Stephenson, Silvain Gerber, Gérard Bailly, and Simon King,
\newblock ``Refining the evaluation of speech synthesis: A summary of the blizzard challenge 2023,''
\newblock {\em Computer Speech \& Language}, vol. 90, pp. 101747, 2025.

\bibitem{muller_speech_2024}
Thomas Muller, Stephane Ragot, Laetitia Gros, Pierrick Philippe, and Pascal Scalart,
\newblock ``Speech quality evaluation of neural audio codecs,''
\newblock in {\em Interspeech 2024}, 2024-09-01, pp. 1760--1764.

\bibitem{naderi_open_2020}
Babak Naderi and Ross Cutler,
\newblock ``An open source implementation of {ITU-T} recommendation {P}.808 with validation,''
\newblock in {\em Interspeech 2020}, 2020-10-25, pp. 2862--2866.

\bibitem{valin_definition_2012}
Jean-Marc Valin, Koen Vos, and Timothy~B. Terriberry,
\newblock ``Definition of the opus audio codec,'' 2012,
\newblock Issue: 6716 Num Pages: 326 Series: Request for Comments Published: {RFC} 6716.

\bibitem{buthe_lace_2023}
Jan Büthe, Jean-Marc Valin, and Ahmed Mustafa,
\newblock ``{LACE}: A light-weight, causal model for enhancing coded speech through adaptive convolutions,''
\newblock in {\em {WASPAA} 2023}. 2023, pp. 1--5, {IEEE}.

\bibitem{buthe_nolace_2024}
Jan Büthe, Ahmed Mustafa, Jean-Marc Valin, Karim Helwani, and Michael~M. Goodwin,
\newblock ``{NOLACE}: Improving low-complexity speech codec enhancement through adaptive temporal shaping,''
\newblock in {\em {ICASSP} 2024}. 2024, pp. 476--480, {IEEE}.

\bibitem{parker_scaling_2025}
Julian Parker, Anton Smirnov, Jordi Pons, {CJ} Carr, Zack Zukowski, Zach Evans, and Xubo Liu,
\newblock ``Scaling transformers for low-bitrate high-quality speech coding,''
\newblock in {\em International Conference on Representation Learning}, Y.~Yue, A.~Garg, N.~Peng, F.~Sha, and R.~Yu, Eds., 2025, vol. 2025, pp. 51997--52021.

\bibitem{xin_bigcodec_2024}
Detai Xin, Xu~Tan, Shinnosuke Takamichi, and Hiroshi Saruwatari,
\newblock ``{BigCodec}: Pushing the limits of low-bitrate neural speech codec,''
\newblock {\em arXiv}, 2024.

\bibitem{gu_esc_2024}
Yuzhe Gu and Enmao Diao,
\newblock ``{ESC}: Efficient speech coding with cross-scale residual vector quantized transformers,''
\newblock in {\em Proceedings of the 2024 Conference on Empirical Methods in Natural Language Processing}, Yaser Al-Onaizan, Mohit Bansal, and Yun-Nung Chen, Eds. 2024-11, pp. 10084--10096, Association for Computational Linguistics.

\bibitem{nc_bm}
Wolfgang Mack, Nezih Topaloglu, Laura Lechler, Ivana Balic, Alexandra Craciun, Mansur Yesilbursa, and Kamil Wojcicki,
\newblock ``Assessing speech quality metrics for evaluation of neural audio codecs under clean speech conditions,''
\newblock {\em arXiv}, 2025.

\end{thebibliography}

\end{document}